\documentclass[
  journal=pasa,
  manuscript=research-paper, 
  year=2025,
  volume=YY,
]{cup-journal}

\usepackage{amsmath}
\usepackage{amssymb,microtype,siunitx,booktabs,hyperref}
\usepackage[table]{xcolor}
\usepackage{tabularx}
\usepackage{booktabs}
\newcommand\arcsec{$^{\prime\prime}\,$}

\title{The Deeper, Wider, Faster programme's first DECam optical data release}

\author{James Freeburn}
\affiliation{Centre for Astrophysics and Supercomputing, Swinburne University of Technology, Hawthorn, 3122, VIC, Australia}
\alsoaffiliation{ARC Centre of Excellence for Gravitational Wave Discovery}
\email[James Freeburn]{jfreeburn@swin.edu.au}

\author{Jeff Cooke}
\affiliation{Centre for Astrophysics and Supercomputing, Swinburne University of Technology, Hawthorn, 3122, VIC, Australia}
\alsoaffiliation{ARC Centre of Excellence for Gravitational Wave Discovery}

\author{Anais M\"oller}
\affiliation{Centre for Astrophysics and Supercomputing, Swinburne University of Technology, Hawthorn, 3122, VIC, Australia}
\alsoaffiliation{ARC Centre of Excellence for Gravitational Wave Discovery}

\author{Jielai Zhang}
\affiliation{Centre for Astrophysics and Supercomputing, Swinburne University of Technology, Hawthorn, 3122, VIC, Australia}
\alsoaffiliation{ARC Centre of Excellence for Gravitational Wave Discovery}

\author{Dougal Dobie}
\affiliation{Sydney Institute for Astronomy, School of Physics, University of Sydney, Sydney, NSW 2006, Australia}
\alsoaffiliation{ARC Centre of Excellence for Gravitational Wave Discovery}

\author{Brent Miszalski}
\affiliation{Australian Astronomical Optics, Faculty of Science and Engineering, Macquarie University, North Ryde, 2113, Australia}

\author{Simon O'Toole}
\affiliation{Australian Astronomical Optics, Faculty of Science and Engineering, Macquarie University, North Ryde, 2113, Australia}

\author{James Tocknell}
\affiliation{Australian Astronomical Optics, Faculty of Science and Engineering, Macquarie University, North Ryde, 2113, Australia}

\author{Sam Huynh}
\affiliation{Australian Astronomical Optics, Faculty of Science and Engineering, Macquarie University, North Ryde, 2113, Australia}

\author{Sara Webb}
\affiliation{Centre for Astrophysics and Supercomputing, Swinburne University of Technology, Hawthorn, 3122, VIC, Australia}
\alsoaffiliation{ARC Centre of Excellence for Gravitational Wave Discovery}

\author{Igor Andreoni}
\affiliation{Department of Physics and Astronomy, University of North Carolina at Chapel Hill, Chapel Hill, NC 27599-3255, USA}

\author{Natasha Van Bemmel}
\affiliation{Centre for Astrophysics and Supercomputing, Swinburne University of Technology, Hawthorn, 3122, VIC, Australia}
\alsoaffiliation{ARC Centre of Excellence for Gravitational Wave Discovery}

\author{Timothy M. C. Abbott}
\affiliation{NOIRLab/MSO/CTIO Casilla 603, La Serena, Chile}

\author{Rebecca Allen}
\affiliation{Centre for Astrophysics and Supercomputing, Swinburne University of Technology, Hawthorn, 3122, VIC, Australia}
\alsoaffiliation{ARC Centre of Excellence for Gravitational Wave Discovery}

\author{Stephanie Bernard}
\affiliation{School of Physics, University of Melbourne, Parkville, 3010, VIC, Australia}

\author{Simon Goode}
\affiliation{School of Physics and Astronomy, Monash University, VIC 3800, Australia}
\alsoaffiliation{ARC Centre of Excellence for Gravitational Wave Discovery}

\author{Sarah Hegarty}
\affiliation{Centre for Astrophysics and Supercomputing, Swinburne University of Technology, Hawthorn, 3122, VIC, Australia}

\author{J. Chuck Horst}
\affiliation{Department of Astronomy, San Diego State University, 5500 Campanile Drive, San Diego, CA 92812}

\author{Cassidy Mihalenko}
\affiliation{School of Natural Sciences, University of Tasmania, Private Bag 37 Hobart, Tasmania, 7001, Australia}
\alsoaffiliation{ARC Centre of Excellence for Gravitational Wave Discovery}

\author{Mark Suhr}
\affiliation{Centre for Astrophysics and Supercomputing, Swinburne University of Technology, Hawthorn, 3122, VIC, Australia}

\doi{10.1017/pasa.2020.32}

\received {dd Mmm YYYY}
\revised  {dd Mmm YYYY}
\accepted {dd Mmm YYYY}
\published{22 September 202X}


\begin{document}

\begin{abstract}
The transient and variable optical sky is relatively poorly characterised on fast ($<$1\,hr) timescales.  With the Dark Energy Camera (DECam), the Deeper, Wider, Faster programme (DWF) probes a unique parameter space with its deep (median of $g\sim22.2$ AB mag), minute-cadence imaging.  In this work, we present DWF's first data release which comprises high cadence photometry extracted from $\sim$12000 images and 166 hours of telescope time.  We present a novel data processing pipeline, \texttt{dwf-postpipe}, developed to identify sources and extract their light curves.  The accuracy of the photometry is assessed by cross-matching to public catalogues.  In addition, we injected a population of synthetic GRB afterglows into a subset of the DWF DECam imaging to compare the efficiency of our pipeline with a standard difference imaging approach.  Both pipelines show performance and reliably recover injected transients with peak magnitudes $g<22$ AB mag with an efficiency of $97.24^{+0.7}_{-1.0}$ percent for \texttt{dwf-postpipe} and $96.14^{+0.9}_{-1.1}$ percent for a difference imaging approach.  However, we find that \texttt{dwf-postpipe} is less likely to recover transients appearing in galaxies that are brighter or comparable in brightness to the transient itself. To demonstrate the power of the data in this release, we conduct a search for uncatalogued variable stars in a single night of DWF DECam imaging and find ten pulsating variables, two eclipsing binaries and one ZZ ceti.  We also conduct a search for variable phenomena in the Chandra Deep Field South, a Rubin deep drilling field, and identify two flares from likely UV ceti type stars.
\end{abstract}

\section{INTRODUCTION}

The fast transient sky, events lasting less than $\sim$1\,hr duration, is not well characterised at optical wavelengths. Radio and high energy facilities are readily capable of detecting short (millisecond-to-hour) duration transients such as Fast Radio Bursts \citep[FRBs;][]{2007Sci...318..777L}, Gamma-ray bursts \citep[GRBs;][]{1973ApJ...182L..85K} and Fast X-ray transients \citep[FXRTs;][]{2022A&A...663A.168Q,2023A&A...675A..44Q}. In optical and infrared however, the size and readout times of charged-coupled devices (CCDs) and wide field format arrays place limits on sky coverage and cadence compared to radio and high energy facilities for faint fast transients.

Despite this, there is a variety of observed optical transients which evolve on sub-hour timescales. The synchrotron afterglows originating from the forward shocks of jets launched from collapsars \citep{1997Natur.387..479G}, neutron star mergers \citep{2005Natur.437..855V,2005Natur.437..859H} and tidal-disruption events \citep{2022Natur.612..430A} exhibit rapidly decaying optical emission. Furthermore, the optical emission associated with a reverse shock can provide even more luminous optical transients decaying by multiple magnitudes over minute timescales \citep[e.g.][]{1999Natur.398..400A,2019MNRAS.489.1820L,2023NatAs...7..843O}. Observations of the Luminous Fast Blue Optical transient (LFBOT) AT2022tsd revealed luminous, minute-timescale flares \citep{2023Natur.623..927H}. This phenomenon is not easily explained with the current understanding of LFBOT progenitors. Supernova shock-breakouts \citep{2010ApJ...725..904N}, occurring on minutes to hours timescales, have a number of known examples \citep{2008Sci...321..223S,2016ApJ...820...23G,2018Natur.554..497B}.  Some of the most commonly observed minute timescale optical transients are stellar flares which arise from magnetic reconnection processes from main-sequence stars \citep{1974ApJS...29....1M}.

There are also a number of theorised fast transients, which to-date, have not been observed.  These include primordial black hole micro-lensing \citep{2005ApJ...634.1103R} and possible optical counterparts to FRBs \citep{2014ApJ...792L..21Y,2019ApJ...878...89Y}, blitzars \citep{2014A&A...562A.137F}, synchrotron masers \citep{2018ApJ...864L..12L,2019MNRAS.485.4091M} and white dwarf accretion-induced collapse \citep{2019ApJ...886..110M}. 

There are a small number of surveys that have probed the fast transient sky in optical wavelengths at shallower depths. Evryscope \citep{2015PASP..127..234L} is a dedicated fast transient survey, which possesses a 18\,400 square degree field of view and a two minute cadence, and has detected a vast population of stellar flares \citep{2018ApJ...860L..30H,2019ApJ...881....9H}. The Transiting Exoplanet Survey Satellite \citep[TESS;][]{2015JATIS...1a4003R} was designed primarily to detect transiting exoplanets from nearby, bright stars. However, in recent years, the TESS data has been appropriated for both targeted \citep[e.g.][]{2024Natur.626..737L,2024ApJ...963...89R,2024ApJ...972..162J,2025MNRAS.537.2362P} and untargeted approaches to the identification of extragalactic fast transients \citep{2025arXiv250216905R}. 

Evryscope and TESS have depths of $\sim$14 and 17 AB mag respectively. Only a small percentage of cosmological transients rise above these limits. Deeper surveys reach further into fast transient population luminosity distributions and probe larger cosmological volumes, allowing for more frequent transient detection. Traditional synoptic surveys such as the Zwicky Transient Facility \citep[ZTF;][]{2019PASP..131a8002B}, The Dark Energy Survey \citep[DES;][]{2016MNRAS.460.1270D}, Pan-STARRS1 \citep{panstarrs} and the upcoming Vera C. Rubin Observatory's Legacy Survey of Space and Time \citep[Rubin;][]{2019ApJ...873..111I} have depths of $\sim$20.5, 23, 24 and 24.5 AB magnitude respectively. However, their main surveys typically have cadences in excess of one day, although some some have conducted specialised, high-cadence experiments \citep[e.g.][]{2013ApJ...779...18B,2019AAS...23325921K,2021MNRAS.505.1254K,2024TNSAN.135....1H}.

The Deeper, Wider, Faster programme \citep[DWF;][]{2019IAUS..339..135A} provides a dedicated deep survey for identifying fast optical transients on minute timescales to $m \sim 23$. Moreover, it is a transient survey which comprises simultaneous observations in radio, millimetre, optical, ultra-violet, X-ray, gamma-ray and high-energy particles. Using the Dark Energy Camera \citep[DECam;][]{2015AJ....150..150F}, mounted on the Victor M. Blanco 4m telescope, DWF probes a new parameter space in the search for fast optical transients with its depth, minute cadence and sky coverage.  Subaru Hypter Suprime-Cam \citep{2002PASJ...54..833M} and KMTNet \citep{2016JKAS...49...37K} have also been used for high cadence optical imaging during DWF runs but these data are not included in this data release.

In this paper, we describe the Deeper, Wider, Faster programme's first data release of the minute-cadence optical data obtained with DECam. In Section \ref{sec:DWF}, we describe the observing strategy utilised with DWF, its uniqueness compared to other surveys and an overview of the DWF observing runs included in this data release. Section \ref{sec:pipeline} describes the data processing pipeline, \texttt{dwf-postpipe}, which is used to produce the data products for this release, and we assess the data quality of the data products in Section \ref{sec:data_quality}. We identify a sample of uncatalogued variable stars using a single night of DWF DECam data and search the Chandra Deep Field South (CDFS), a Rubin deep-drilling field, for variable phenomena and identify two flare stars in Section \ref{sec:variables}. Section \ref{sec:discussion} discusses classification of the identified variables, potential applications of this dataset, and future facilities that will complement it.

\section{DATA}\label{sec:DWF}

\subsection{The minute-timescale optical sky}

\begin{figure*}
    \includegraphics[width=0.65\textwidth]{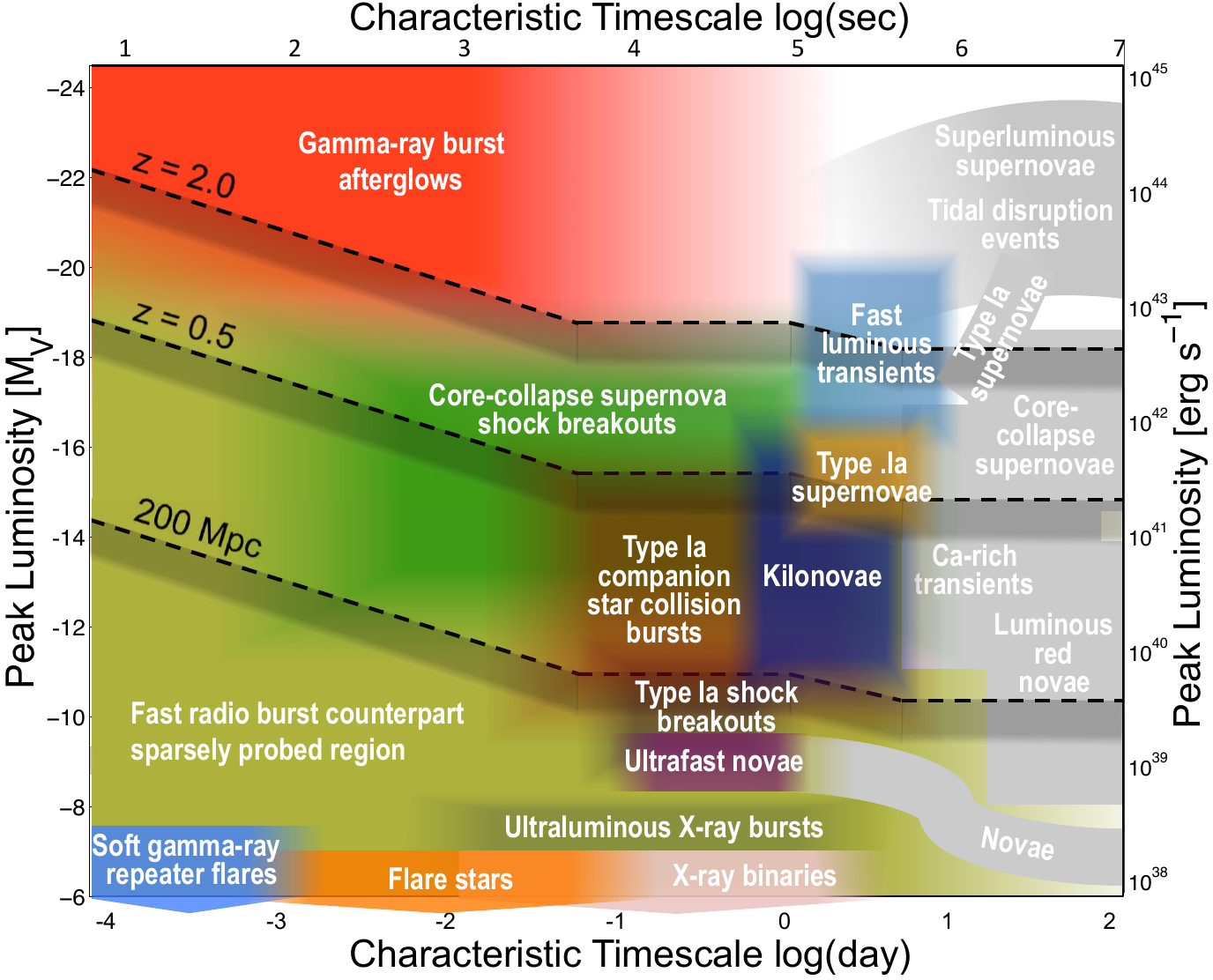}
    \caption{Approximate timescales and luminosities of known and theorised optical transients. The peak luminosities of Soft-gamma-ray repeater flares, flare stars and X-ray binaries extend lower than the axes shown on the plot for readability. Adapted from Cooke et al., in preparation.}
    \label{fig:FT_timescales}
\end{figure*}

Transients with characteristic timescales of $\gtrsim10^1$\,d, which includes most classes of supernova (see Fig. \ref{fig:FT_timescales}), are readily detected by traditional optical surveys, which typically revisit the same fields on daily to weekly cadences \citep{2020ApJ...904...35P}. In contrast, the minute-timescale sky is dominated by frequent Galactic events, such as flare stars \citep{1974ApJS...29....1M, 1991ARA&A..29..275H}, while extragalactic phenomena such as Gamma-ray burst afterglows \citep[e.g.][]{2022ApJ...938...85H,2024MNRAS.531.4836F} and theorised counterparts to fast radio bursts are far rarer \citep[e.g.][]{2018RNAAS...2...30P,2019ApJ...881...30T,2021A&A...653A.119N,2025MNRAS.538.1800H}.  DWF probes the deep, minute-timescale optical sky in an effort to identify these rare classes.

DECam has been the dominant wide-field imager used in most DWF coordinated observing runs to detect minute-timescale optical transients and identify counterparts to multi-wavelength transients. Our team reduce and analyse the data in real-time with the \texttt{Mary} pipeline \citep{2017PASA...34...37A}, aided by a convolutional neural network \citep{2022MNRAS.513.1742G} and data visualisation tools \citep[e.g.][Hegarty, in prep.]{2017PASA...34...23M} to trigger rapid response target-of-opportunity observations. 

To-date, the DECam dataset of DWF has been used to study the population of extragalactic fast transients \citep{2020MNRAS.491.5852A}, stellar flares \citep{2021MNRAS.506.2089W,2025arXiv250719584C} and orphan afterglows \citep{2024MNRAS.531.4836F}, among other things.  There have also been multiple, novel classification methods developed for DWF DECam light curves for different purposes \citep{2020MNRAS.498.3077W,2022AJ....163...95S,2024MNRAS.531.4836F}.

\subsection{Observations included in this data release}

For this data release, we include observations taken via the main observing strategy of DWF which involves taking continuous 20\,s $g$-band exposures for periods ranging from $\sim$0.5--3 hours per field. With DECam's $\sim$30 second overhead between exposures, this corresponds to a cadence of $\sim$50 seconds. On a given night, 2--5 fields are targeted and the observations are repeated for 6 consecutive nights during each observing run. As a result, each run produces 6 sets of nightly high cadence data for each field, separated by $\sim$21--23.5\,hr. For the remainder of this work, we refer to a given set of high-cadence observations for a specific field and night by `field-night.'  

To-date, there have been 14 DWF coordinated operational runs, denoted O1 through O14, from December 2015 to January 2025 and two Pilot runs in January and February 2015. During observing runs DWF O4, O8 and O11, the fast-cadenced optical observations were performed with Subaru Hyper Suprime-Cam and KMTNet, respectively. As this work reports specifically on the DWF DECam data, we exclude these runs. DWF O9 used an alternate strategy with dithered exposures and a larger number of fields tailored to detect kilonovae and had fewer exposures in each field per night which served as a pilot run for the Kilonova and Transients Programme \citep[KNTraP;][]{2025MNRAS.537.3332V}. In addition, the two DWF pilot runs employed a dithering strategy, that is not compatible with the photometric pipeline in this data release. Finally, in DWF O13 and O14, three fields in the Large Magellanic Cloud were targeted. Due to the high density of sources, these fields are confusion limited by our processing and are therefore not included in this data release.  As a result, we use only data from the remaining nine DWF runs.  Occasionally, time was taken out of the main observing strategy for rapid target-of-opportunity observations of multi-wavelength transients \cite[e.g.,][]{2025MNRAS.537.2061F}. These observations depart from the primary observing strategy of DWF and include longer exposure times, dithering and multiple filters. We therefore do not include these observations in this data release.  Full details of DWF programme including all multi-wavelength facilities and operational runs will be described in Cooke et. al (in prep.).

\begin{figure}
    \includegraphics[width=\columnwidth]{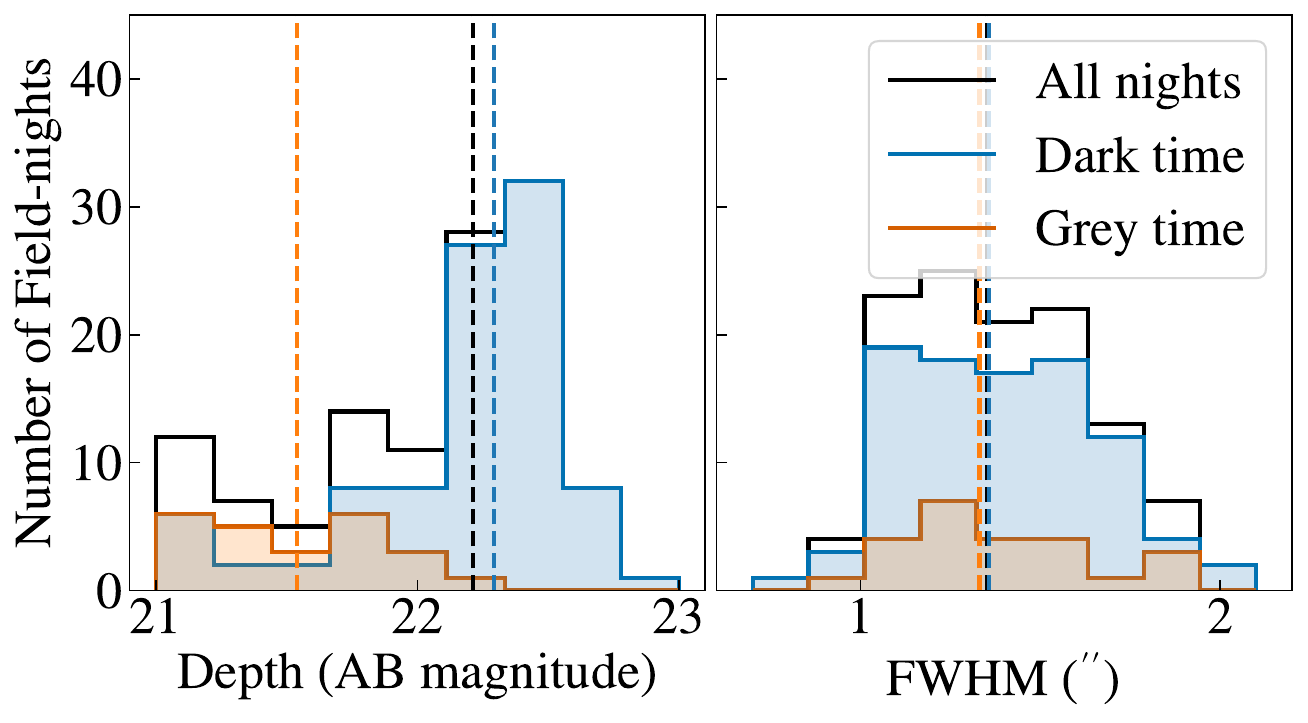}
    \caption{Histograms of the median $g$-band limiting magnitudes (left) and median seeing FWHM (right) for the field-nights included in this data release. The $g$-band depth and seeing FWHM are affected by the need for DECam observations to occur up to relatively high airmass ($\sim$1.3--2.0) in order to enable simultaneous observations of each field by telescopes in Chile, Australia and other parts of the world. Dark time is defined as field-night observations that begin and end with a moon below the horizon and grey time is defined as those that begin or end with a moon altitude $>0^{\circ}$. The median values for each distribution are indicated by the vertical dashed lines.}
    \label{fig:histograms}
\end{figure}

The remaining DECam observations are summarised in Table \ref{tab:fields}, comprising $\sim12000$ images and $166$\,hr of observing time over 112 field-nights.  The median limiting magnitude and full width at half maximum (FWHM) value for each exposure across these field-nights is $g\sim22.2$ and 1.35\arcsec respectively. DWF is scheduled for 6 consecutive nights during dark time, with O2 being conducted across 13 nights. Consequently, 80\% of DECam observing time is conducted with the moon below the horizon. The remaining 20\% has a range of lunar illumination fractions. We remove nights where the median $>5\sigma$ limiting magnitude is $<21$ AB mag.   The resulting distributions are shown in Fig. \ref{fig:histograms}. Dark time observations have a comparatively deeper median limiting magnitude of $g=22.3$ (5$\sigma$) than grey time, $g=21.5$ (5$\sigma$). The median seeing value of 1.35\arcsec remains consistent between bright and dark time observations. Given that 80\% of DWF DECam observations were conducted during dark time, we obtain an overall median 5$\sigma$ limiting magnitude of $g=22.2$. 

\begin{figure*}
    \includegraphics[width=\textwidth]{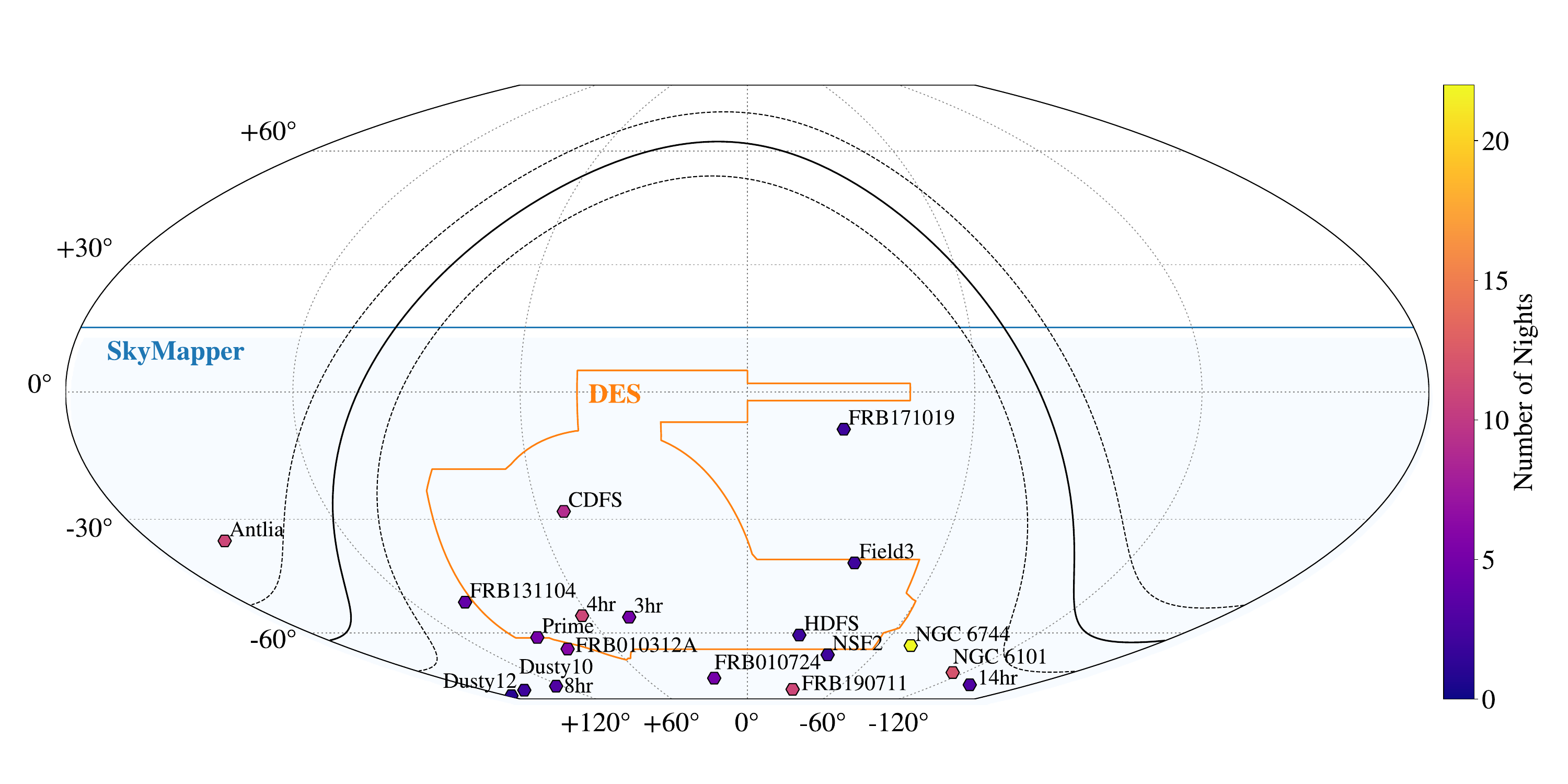}
    \caption{Sky locations of the DWF fields presented in this data release. A solid black line marks the Galactic plane and the dotted black lines denote $\pm10^{\circ}$ from the plane. The number of nights each field has been observed is indicated by the colourbar on the right.  The footprints of the SkyMapper Southern Survey DR4 \citep[SMSS DR3;][]{2024PASA...41...61O} and DES \citep{2016MNRAS.460.1270D} are shown in blue and orange, respectively.}
    \label{fig:fields}
\end{figure*}

DWF is mainly designed to detect extragalactic fast transients and the fields are selected based on a number of criteria. The DECam observations probe a large cosmological volume and, as such, extragalactic transients can occur in any random pointing. However, fields are selected at high Galactic latitude, when possible, to minimise Galactic extinction for the optical and some shorter-wavelength observations. As searching for counterparts to fast radio bursts (FRBs) has been a main aim of DWF since its inception in 2014, some DWF fields target the sky location of repeating FRBs in an effort to detect a repeat burst with simultaneous, multi-wavelength coverage, while searching the wide-fields for new FRBs. Other considerations include targeting known galaxy clusters and legacy fields where there is a large amount of multi-wavelength data and spectroscopic characterisation of galaxies, and fields with previous DWF observations for temporal monitoring and deep image templates. Finally, field choice is constrained to fields that are simultaneously observable from the locations of each major multi-wavelength facility (often in Chile and Australia) participating in a given DWF observing run and the scheduled nights. The sky locations of the fields are shown in Fig. \ref{fig:fields} along with the number of nights on which they were observed.

\begin{table}
 \caption{Fields and night coverage for this data release.}
 \label{tab:fields}
 \begin{tabular*}{\columnwidth}{@{}l@{\hspace*{7pt}}l@{\hspace*{7pt}}l@{\hspace*{7pt}}l@{\hspace*{7pt}}l@{\hspace*{15pt}}}
  \hline
  Field & Coordinates & Gal. Latitude & Runs & Nights\\
  & & $^\circ$ & & \\
  \hline
  FRB010724 & 01:18:06 -75:12:19 & -41.80 & Dec 2015 (O1) & 5\\
  3hr & 03:00:00 -55:25:00 & -53.43 & Dec 2015 (O1) & 5\\
  CDFS & 03:30:24 -28:06:00 & -54.93 & Dec 2015 (O1), & 9\\
    & & & Dec 2019 (O7), & \\
    & & & Sep 2021 (O10), & \\
    & & & Sep 2022 (O12) & \\
  4hr & 04:10:00 -55:00:00 & -44.76 & Dec 2015 (O1), & 11\\
    & & & Jan 2025 (O14) & \\
  Prime & 05:55:07 -61:21:00 & -30.26 & Feb 2017 (O3) & 5\\
  FRB\,131104 & 06:44:00 -51:16:00 & -21.95 & Feb 2017 (O3) & 4\\
  8hr & 08:16:00 -78:45:00 & -22.62 & Jun 2018 (O5) & 3\\
  Dusty 10 & 10:12:00 -80:50:00 & -19.96 & Jun 2018 (O5) & 2\\
  Antlia & 10:30:00 -35:20:00 & 19.17 & Feb 2017 (O3), & 11\\ 
   & & &  Jun 2018 (O5) & \\
   & & &  Jan 2024 (O13) & \\
  Dusty 12 & 11:46:00 -84:33:00 & -21.89 & Jul 2016 (O2) & 1\\ 
  14hr & 14:34:00 -78:06:00 & -16.30 & Jul 2016 (O2) & 3\\
  NGC 6101 & 16:26:00 -73:00:00 & -16.37 & Jul 2016 (O2),  & 12\\
   & & & Aug 2016 (O2), & \\
   & & & Jun 2019 (O6) & \\
  NGC 6744 & 19:09:46 -63:51:27 & -26.15 & Jul 2016 (O2),  & 22\\
   & & & Aug 2016 (O2), & \\
   & & & Jun 2019 (O6), & \\
   & & & Sep 2021 (O10), & \\
   & & & Sep 2022 (O12) & \\
  Field 3 & 21:00:00 -42:48:00 & -49.41 & Jun 2019 (O6) & 2\\
  NSF2 & 21:28:00 -66:48:00 & -39.82 & Jul 2016 (O2) & 2\\
  FRB190711 & 21:57:41 -80:21:29 & -33.90 & Sep 2021 (O10), & 11\\
   & & & Sep 2022 (O12) & \\
  FRB\,171019 & 22:17:31 -08:39:32 & -49.24 & Dec 2019 (O7), & 2\\
  HDFS & 22:33:26 -60:38:09 & -49.22 & Sep 2021 (O10) & 2\\  
  \hline
  Total &  &  & & 112\\
  \hline
 \end{tabular*}
\end{table}

\section{DWF-POSTPIPE}\label{sec:pipeline}

The real-time data processing pipeline, \texttt{Mary}, with a difference-imaging approach, enables the rapid identification of transients for triggering during DWF runs.  In this work, we design a custom pipeline, \texttt{dwf-postpipe}, to be used in post-run to generate an archival dataset.  \texttt{dwf-postpipe} does not utilise difference imaging. A light curve is produced for every object in the field, regardless of whether there is an observed change in brightness. This aids in the discovery of variable stars as, with a difference imaging approach, their detectability would depend on their phase when the template image was taken.  It also reduces the number of artefacts due to poor alignment and image quality matching between templates and science images.  We distinguish the three pipelines used for processing DWF DECam data mentioned in this work: \texttt{Mary} the real-time data processing pipeline; \texttt{dwf-postpipe}, the post-run pipeline described below; and \texttt{photpipe} \citep{2005ApJ...634.1103R,2014ApJ...795...44R}, which is a separate pipeline used for comparative analysis in Section \ref{sec:fakes}.

\begin{figure*}
    \centering
    \includegraphics[width=0.8\textwidth]{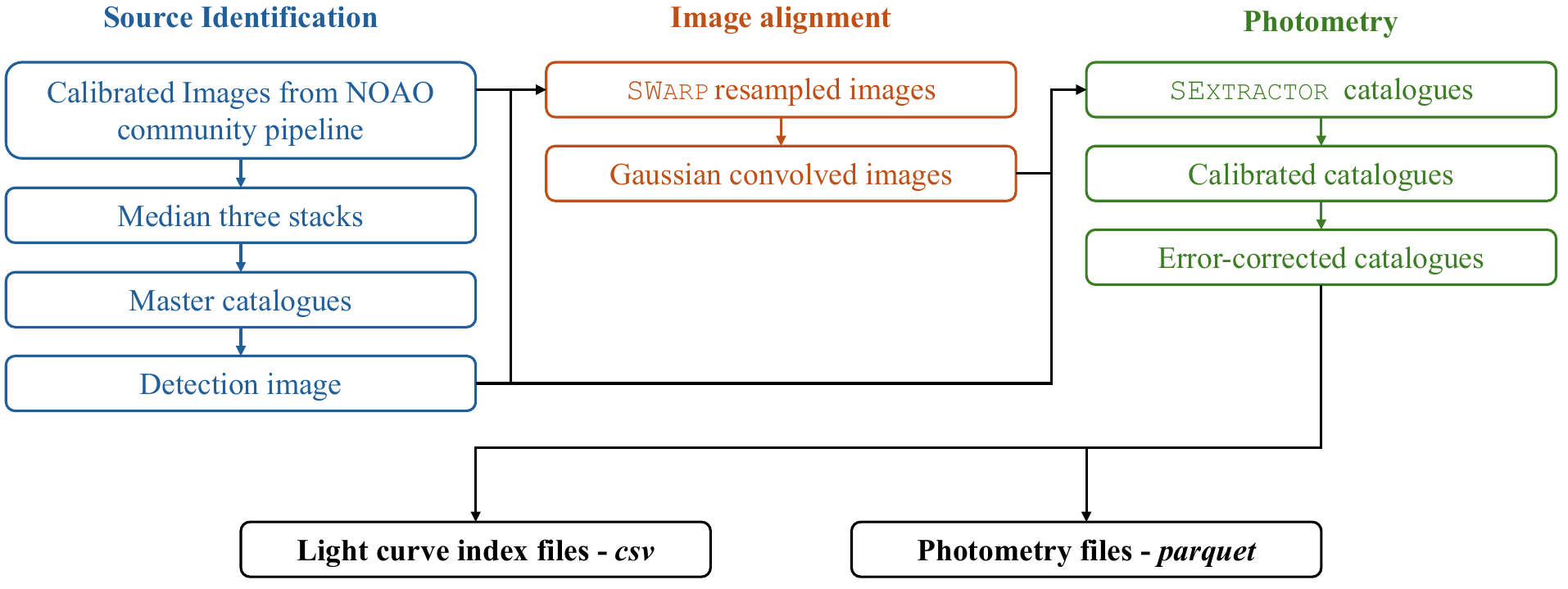}
    \caption{Schematic diagram of \texttt{dwf-postpipe} used in this data release. This methodology is applied to each of the 112 field-nights shown in Table \ref{tab:fields}.}
    \label{fig:pipeline}
\end{figure*}

We designed \texttt{dwf-postpipe} primarily to produce intra-night, minute-cadence light curves reflecting the DWF DECam observing strategy and main goal of identifying transients evolving on minute timescales.  The pipeline begins with the calibrated images from The DECam Community Pipeline \citep{2014ASPC..485..379V}, which applies 
corrections to cross-talk, overscan, fringing in addition to flat-fielding, bias-correction and an astrometric solution. It also performs cosmic ray and saturation rejection and background subtraction.  It is a robust pipeline which produces good quality images, suitable for photometry. \texttt{dwf-postpipe} comprises three steps: Source identification, image matching and photometry. Below, we explain each step in detail and show a graphical representation of the pipeline in Fig. \ref{fig:pipeline}. The data products from \texttt{dwf-postpipe} have already been utilised for a search for GRB afterglows \citep{2024MNRAS.531.4836F} and assessing simultaneous optical and radio variability of radio transients found with the Australia Square Kilometer Array Pathfinder \citep{2023MNRAS.519.4684D}.

\subsection{Source identification}

\texttt{dwf-postpipe} is designed to generate complete light curves for each source detected throughout the night for a given DECam pointing. In this effort, we want to ensure transient sources that rise above or fall below the detection threshold throughout the night are identified and measured. This requires two components; a `master catalogue' of all of the sources detected throughout the night, transient or not, and a method to perform forced photometry at the location of each of these sources for every image taken throughout the night. 

We use \texttt{SExtractor} \citep{sextractor}, a code that detects, deblends, measures and classifies sources in astronomical imaging.  For source identification, we use \texttt{SExtractor}’s built-in background modelling algorithm which estimates the image background by calculating values from a mesh grid and interpolating across the image using a cubic spline.  For photometry, background values for each source are estimated by using \texttt{SExtractor}’s \texttt{LOCAL} method which is based on a rectangular annulus with a width of 24 pixels.  \texttt{SExtractor} does not include an in-built method for performing forced photometry directly from a catalogue. However, \texttt{SExtractor} includes the capability to run in `double image' mode that identifies sources from a `detection image' and then measures photometry, classification and deblending on a `measurement image.' We use this method for this pipeline.  

To minimise the contamination of electronic artefacts in the master catalogue, we initially conduct a three-image rolling median coaddition. With this method, sources appearing in a single image in the night are not included in the master catalogue and do not have an associated light curve. \texttt{SExtractor} generates catalogues for each of these stacked images.  The catalogues are then compiled into a master catalogue by cross-matching sources within one arcsecond of each successive catalogue.  If a source in a successive catalogue is not cross-matched, has $>5\sigma$ detection and a FWHM within 1$\sigma$ of the median, point-source value, it is appended to the master catalogue.  This is to ensure only genuine transients are added to the master catalogue rather than electronic artefacts like cross-talk.

We select the detection image in a given night as the image with the highest point source seeing FWHM value to ensure the PSF is matched properly for each image in a field-night. This process is explained in more detail in Section \ref{sec:matching}. Sources that are in the master catalogue but do not appear in the detection image are artificially added to the detection image with a Moffat profile and a FWHM in accordance with the median, point-source value of the detection image. In dual-image mode, therefore, each source in the master catalogue is identified as a source in the detection image.

\subsection{PSF matching}\label{sec:matching}

Once sources are added to the detection image, to ensure pixel-for-pixel alignment between images, we resample the each image for given field and night to the detection image using \texttt{Swarp} \citep{SWarp}.  

Variability in atmospheric seeing between images can result in flux from neighbouring sources contaminating photometry to varying degrees in different images.  This induces time-correlated noise into the photometry of blended sources, which can easily be mistaken for astrophysical variability.  To minimise this effect, we match each image's point spread function (PSF) to the detection image with a Gaussian convolution using \texttt{hotpants} \citep{hotpants}.  Adopting the image with the highest FWHM value as the detection image ensures that the PSF of every other image of the night is degraded to match the detection image.

\subsection{Photometry}

Photometry is the third and final step of the pipeline. We use \texttt{SExtractor} in double-image mode with the resampled detection image to detect sources and the resampled, convolved images from the previous step for measurement. Because transient sources have been added to the detection image, this will be the complete set of sources for the night. \textit{SkyMapper}'s \citep{2024PASA...41...61O} fourth data release (DR4) includes coverage of all the DWF fields included in this data release.  We therefore use \textit{SkyMapper} DR4 to calculate magnitude zeropoints for the instrumental photometry for each CCD for each exposure to ensure a uniform calibration.  

\texttt{SExtractor} estimates the error of photometric measurement assuming Poisson statistics between uncorrelated pixels. With the Gaussian convolution induced by \texttt{hotpants}, neighbouring pixels become correlated, resulting in an underestimation of the flux error. To correct for this, we adopt an empirical approach to correct for the reduced effective number of statistically independent pixels.  Due to the high cadence of the DWF data, successive images can be assumed to be similar in depth.  Therefore, the median absolute differences in flux measurements of non-variable sources in successive exposures can be used to estimate the effective error associated the flux measurements. With this assumption, we can calculate estimate the true uncertainty by binning sources into magnitude bins and calculating the median difference in magnitude between a given exposure and the next. This difference is then compared to the median error in each of these bins. The flux error is then multiplied by a factor to resolve the disparity between the median error and median magnitude difference in each bin. This calculation is described in detail in Appendix \ref{sec:error_correction}.

\subsection{Data products}\label{sec:data_products}

The above steps result in each source having a complete light curve over the course of a given night. The light curves from each night are combined for each field by cross-matching those within a one arcsecond radius and labelling each with a object identifier consistent between nights. With this object identifier, candidates can be monitored over DWF's $\sim$10\,yr dataset.

An `Apache Parquet' file is produced containing each photometric measurement for each source in each exposure over the entire DWF dataset which is called \texttt{lightcurve}. In addition, there is a light curve index file, \texttt{mastercatalogue}, containing a row for each source appearing over the course of a single night. Each row represents a nightly light curve that can be queried from the photometric catalogue. The rows contain light curve metadata, such that interesting light curves or objects can be identified without opening the large, photometric parquet file.  These files will be supplied in this release in addition to the reduced images from the DECam community pipeline. The schema for these catalogues are shown in Tables \ref{tab:mastercatalogue} and \ref{tab:lightcurve} in \ref{sec:schema}.

In Table \ref{tab:all_candidates}, we summarise the quantity of candidates observed in this data release.  To quantify variability we use the Von-Neumann statistic \citep{VN_statistic}, 
\begin{equation}
    \eta = \dfrac{\sum_{i=1}^{N-1} (m_{i+1}-m_i)^2/(N-1)}{\sum_{i=1}^{N}(m_{i+1}-\Bar{m})^2/(N-1)}
    \label{eq:vn}
\end{equation}
which describes how much of the variance is explained by intrinsic variability versus random noise. $m_i$ describes the series of measurements in the light curve and N is the total number of measurements in the light curve.  This statistic has been shown to be reliable in identifying variable astrophysical phenomena \citep{vn_comparison}.  Of the 4,862,636 unique objects observed in all fields in this data release, we find that 92\% of them do not exhibit significant variability ($\eta^{-1} < 1$).  The remaining 8\% include astrophysical transients and variables in addition to asteroids and electronic artefacts.  We cross-match 6,927 of these objects with existing variable catalogues; \textit{Gaia} DR3 \citep{Gaia_DR3}, the ASAS-SN catalogue of variable stars \citep{ASAS-SN_Variables} and the catalogue for RR Lyrae variable stars in DES Y6 \citep{2021ApJ...911..109S} which is shown in the known variables column in Table \ref{tab:all_candidates}.

\begin{table}
 \caption{Quantity of unique objects observed in this data release.}
 \label{tab:all_candidates}
 \begin{tabular}{llll}
    \hline
     Field & Unique objects & Objects with $\eta^{-1}>1$ & Known variables \\
    \hline
    FRB010724 & 423,392 & 44,377 & 522 \\
    3hr & 287,127 & 10,248 & 210 \\
    CDFS & 269,714 & 6,776 & 705 \\
    4hr & 370,886 & 20,749 & 238 \\
    Prime & 279,211 & 9,752 & 315 \\
    FRB131104 & 175,979 & 5,987 & 361 \\
    8hr & 136,299 & 6,173 & 316 \\
    Dusty10 & 168,121 & 3,248 & 372 \\
    Antlia & 345,008 & 13,526 & 513 \\
    Dusty12 & 130,120 & 11,430 & 333 \\
    14hr & 340,002 & 44,858 & 597 \\
    NGC 6101 & 499,261 & 83,556 & 832 \\
    NGC 6744 & 578,019 & 87,437 & 503 \\
    Field3 & 109,565 & 1,241 & 319 \\
    NSF2 & 658,64 & 6,818 & 245 \\
    FRB190711 & 544,302 & 44,043 & 230 \\
    FRB171019 & 517,68 & 892 & 111 \\
    HDFS & 87,998 & 1,820 & 205 \\
    \hline
    Total & 4,862,636 & 402,931 & 6,927 \\
    \hline
    \end{tabular}
\end{table}

\subsection{AAO Data Central}
The  calibrated images from the DECam community pipeline (individual epochs) and \texttt{dwf-postpipe} light curves are hosted by the AAO Data Central science platform.\footnote{\url{https://datacentral.org.au}}$^{,}$\footnote{\url{https://doi.org/10.57891/gy62-bj96}} The AAO Data Central science platform provides a wide range of services to engage with DWF data products. The images and light curves are primarily accessible, respectively, via Data Central's Simple Image Access (SIA) and Table Access Protocol (TAP) services. These services enable both programmatic data access e.g. via Python or interactive data access e.g. via \texttt{topcat} \citep{2005ASPC..347...29T}. Individual objects of interest may also be visualised using Data Central's Single Object Viewer that facilitates interactive display of the images, light curves and other metadata in one place. Since these services are under active development, we refer to the latest documentation on the Data Central science platform for the latest information.

\section{DATA QUALITY ANALYSIS}\label{sec:data_quality}

\subsection{Photometric calibration}

The DECam Local Volume Exploration Survey \citep[DELVE;][]{2021ApJS..256....2D} provides $g$-band, DECam imaging coverage of all DWF fields in this data release. With a 5$\sigma$ limiting magnitude of $g\sim23.5$, it provides a useful direct comparison for the photometric pipeline presented in Section \ref{sec:pipeline}. We find that for each DWF field, there is a consistent offset from DELVE and no dependence with magnitude. We demonstrate this in Table \ref{tab:zeropoints}. The $\sim$0.15 mag offset is due to our calibration with SMSS DR4, which has a systematic offset with DELVE.  The systematic offset likely results from SMSS DR4's use of spectroscopic calibrator stars to calculate zero-points in contrast to DELVE's use of the ATLAS Refcat2 catalogue \citep{2018ApJ...867..105T}.

\begin{table*}
 \centering
 \caption{Data quality statistics of each DWF field included in this data release. Magnitude offsets of the median magnitude for each source in each DWF field are cross matched with DELVE photometry. We also show the density of sources in each field with the median number of sources found in each $9^\prime\times18^\prime$ CCD.}
 \label{tab:zeropoints}
 \begin{tabular*}{0.68\textwidth}{@{}l@{\hspace*{17pt}}l@{\hspace*{17pt}}l@{\hspace*{17pt}}l@{\hspace*{17pt}}l@{\hspace*{17pt}}}
  \hline
  Field & DELVE Offset & Median depth & Median FWHM & Median Sources per CCD\\
        & Magnitude    & AB magnitude & \arcsec & \\ 
  \hline
  FRB010724 & $-0.173 \pm 0.027$ & 21.5 & 1.5 & 2,830\\
  3hr & $-0.140 \pm 0.022$ & 21.5 & 1.5 & 1,039\\
  CDFS & $-0.157 \pm 0.031$ & 22.3 & 1.2 & 1,185\\
  4hr & $-0.151 \pm 0.024$ & 22.1 & 1.2 & 1,697\\
  Prime & $-0.159 \pm 0.021$ & 22.3 & 1.3 & 2,602\\
  FRB\,131104 & $-0.158 \pm 0.015$ & 22.4 & 1.4 & 1,445\\
  8hr & $-0.166 \pm 0.013$  & 21.8 & 1.5 & 1,525\\
  Dusty 10 & $-0.162 \pm 0.013$ & 22.0 & 1.3 & 1,801\\
  Antlia & $-0.159 \pm 0.013$ & 22.4 & 1.1 & 2,917\\ 
  Dusty 12 & $-0.168 \pm 0.018$ & 22.6 & 1.3 & 2,202\\ 
  14hr & $-0.160 \pm 0.017$ & 21.2 & 1.4 & 2,969\\
  NGC 6101 & $-0.172 \pm 0.013$ & 22.3 & 1.6 & 3,520\\
  NGC 6744 & $-0.139 \pm 0.013$ & 22.2 & 1.5 & 2,295\\
  Field 3 & $-0.147 \pm 0.022$ & 22.4 & 1.7 & 1,305\\
  NSF2 & $-0.137 \pm 0.017$ & 21.2 & 1.6 & 871\\
  FRB190711 & $-0.158 \pm 0.019$ & 22.1 & 1.5 & 493\\
  FRB171019 & $-0.146 \pm 0.022$ & 21.1 & 1.4 & 2,044\\
  HDFS & $-0.148 \pm 0.017$ & 22.0 & 1.4 & 1,311\\ 
  \hline
 \end{tabular*}
\end{table*}

\subsection{Injection of synthetic transients}\label{sec:fakes}

To assess \texttt{dwf-postpipe}'s performance, we compare the results of our pipeline to \texttt{photpipe} \citep{2005ApJ...634.1103R,2014ApJ...795...44R}, a popular difference imaging pipeline for identifying transient sources.  

For the difference image analysis using \texttt{photpipe}, templates were selected based on the availability of images taken during DWF operational runs with low seeing values.  We coadd 5 NGC\,6101 images and 10 CDFS images which have $5\sigma$ depths of $g=23.2$ and 23.6 AB mag respectively and FWHM values of 1.22\arcsec and 1.27\arcsec respectively.

We inject synthetic transients, hereafter called `fakes', into the DWF data with \texttt{afterglowpy} \citep{2020ApJ...896..166R} using the procedure described in \cite{2024MNRAS.531.4836F}. These events are well suited to test our pipeline due to their fast evolution (i.e., evolution in a single field-night observation). The fakes are injected directly into each image with a Moffat profile matched to the seeing FWHM of that image and placed both coincident with galaxies and randomly throughout the field.

The fakes were injected into five CCDs randomly selected from two fields; CDFS which is a sparse, high Galactic latitude field and NGC 6101 which is a comparatively crowded, low Galactic latitude field.  These fields were chosen to be representative of the range of crowdedness for all fields in this dataset.  Fakes were injected for two nights from these two fields, 10 September 2021 and 30 August 2022 for CDFS and 4 August 2016 and 24 June 2019 for NGC 6101. These nights have seeing FWHM values representative of the entire dataset, spanning 1.1\arcsec to 2\arcsec. A total of 1463 fakes were injected into the images across these four field-nights.

\begin{figure}
    \centering
    \includegraphics[width=\columnwidth]{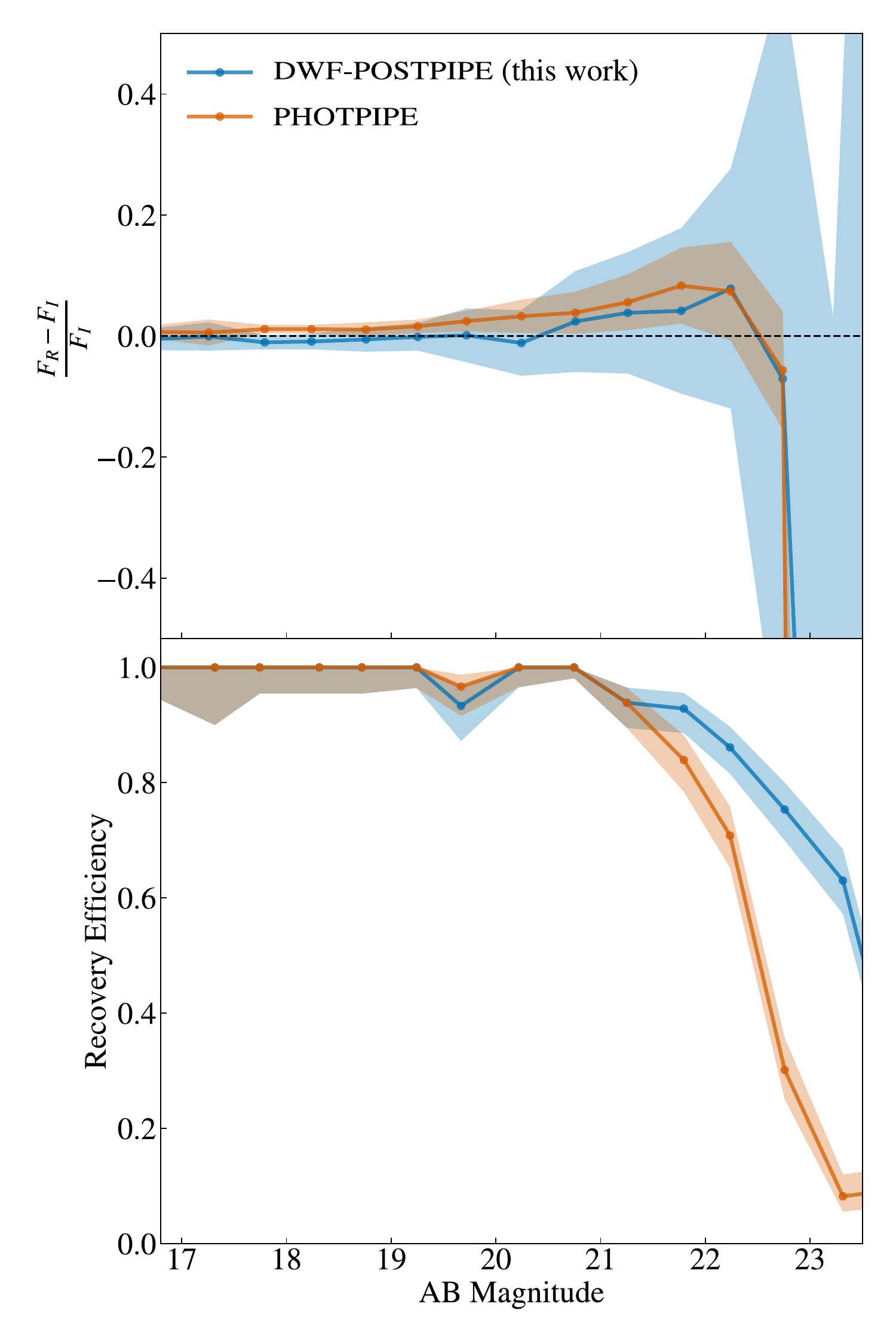}
    \caption{Top panel: Relative difference between the injected flux, $F_I$ and recovered flux, $F_R$, with injected AB magnitude. These values are calculated from each data point independently from each injected `fake' source. Bottom panel: Efficiency at which injected fakes with the peak magnitude are recovered as detections using the data processing pipelines \texttt{photpipe} and \texttt{dwf-postpipe}. Both panels show only the fakes that were injected randomly throughout the field, excluding those that were injected onto galaxies.}
    \label{fig:fluxdiff}
\end{figure}

Fig. \ref{fig:fluxdiff} shows the relative difference between injected and recovered flux for both \texttt{photpipe} and \texttt{dwf-postpipe}. The results from this plot are for fakes injected randomly throughout the field, not those injected onto galaxies. The results are consistent with the limiting magnitude distribution in Fig. \ref{fig:histograms}, with a divergence from injected values at $g\sim22$ AB mag.  Fakes with observed peak magnitudes $g<22$ AB mag are recovered with an efficiency of $97.24^{+0.7}_{-1.0}$ percent for \texttt{dwf-postpipe} and $96.14^{+0.9}_{-1.1}$ percent for \texttt{photpipe}.  There is a sharp drop in recovery efficiency for fakes peaking at $g>22$ AB mag for both pipelines. However, we find that this drop is shallower for \texttt{dwf-postpipe}, resulting in more subthreshold detections.  \texttt{dwf-postpipe} recovers transients $g>22$ with an efficiency of $63.9^{+2.7}_{-2.8}$ percent compared to \texttt{photpipe}'s $29.3^{+2.7}_{-2.6}$ percent. Given the low signal-to-noise ratio ($<5\sigma$) of these detections, it may be difficult to use them to assess a given transient's nature.  In image subtraction, background RMS is increased in quadrature with the template background RMS.  Given the depth of our selected templates (23.2 and 23.6 AB mag for NGC\,6101 and CDFS respectively), we expect this to result in a difference depth $\sim0.1$ magnitudes shallower than the science images.  Therefore, the divergence in recovery efficiency for \texttt{dwf-postpipe} and \texttt{photpipe} can be likely attributed to the loss of sensitivity imparted by image subtraction.  We conclude that the two pipelines are roughly equivalent in their source extraction performance.

However, for sources injected into galaxies, noise induced by the galaxy emission reduces the signal-to-noise ratio for \texttt{dwf-postpipe} compared to \texttt{photpipe}.  As \texttt{dwf-postpipe} does not employ a difference imaging approach, fast transients are typically identified via the variability evident in their light curves.  To quantify variability, we use the von Neumann statistic \citep{VN_statistic} which is explained in Section \ref{sec:data_products}.

\begin{figure}
    \includegraphics[width=\columnwidth]{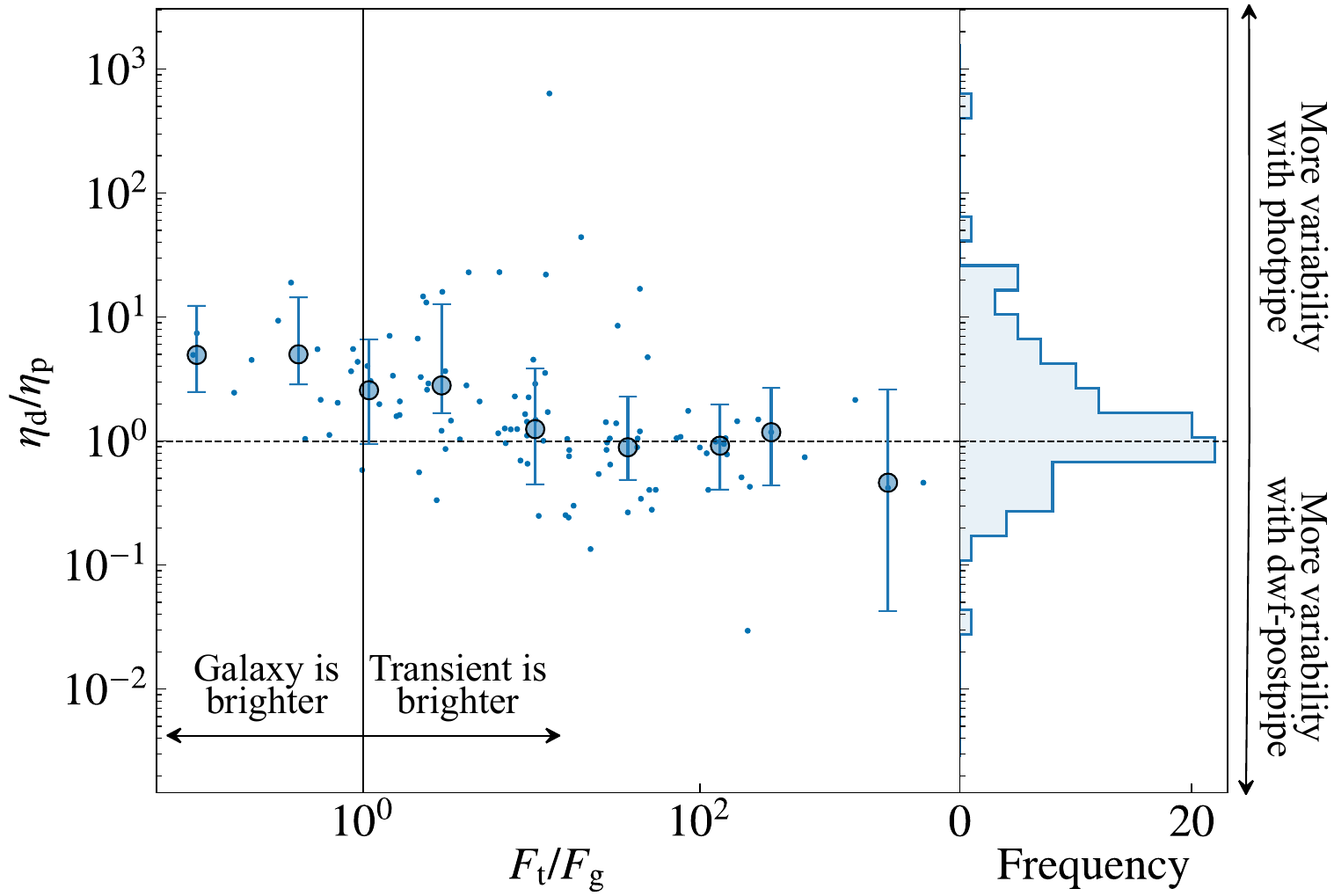}
    \caption{Comparative variability of injected fakes extracted with \texttt{dwf-postpipe} compared with \texttt{photpipe}. The left-hand panel's y-axis is the ratio of the von Neumann statistic from a given light curve extracted with \texttt{dwf-postpipe}.$\eta_\mathrm{d}$ to the same light curve extracted with \texttt{photpipe}, $\eta_\mathrm{p}$. Its x-axis is the ratio the peak flux density of the injected transient, $F_\mathrm{t}$, to the flux density of the galaxy it has been placed onto, $F_\mathrm{g}$. The small points denote individual injected fakes, while the larger points are the median, binned values with error bars denoting the standard deviations for each bin. For injected fakes where the galaxy light is comparable in brightness to its peak brightness or lower ($F_\mathrm{t}$/$F_\mathrm{g} < 10$), variability is more evident in the \texttt{photpipe} light curve compared to the \texttt{dwf-postpipe} light curve ($\eta_\mathrm{d}$/$\eta_\mathrm{p} > 1$). The right-hand panel shows the histogram of $\eta_\mathrm{d}$/$\eta_\mathrm{p}$ values across the entire plot. We see an excess of high values of $\eta_\mathrm{d}$/$\eta_\mathrm{p}$, which are light curves that display more variability when extracted with \texttt{photpipe} compared to \texttt{dwf-postpipe}.}
    \label{fig:galaxy_injections}
\end{figure}

A subset of fakes were injected at the locations of galaxies in the field, identified from  \texttt{sextractor}'s SPREAD\_MODEL parameter, a star/galaxy classifier based on PSF models. Sources with SPREAD\_MODEL $> 0.01$ were assumed to be galaxies.  We demonstrate in Fig. \ref{fig:galaxy_injections} that, for transients injected into galaxies, the $\eta^{-1}$ values of light curves extracted with \texttt{dwf-postpipe} is lower than those extracted with \texttt{photpipe} for the same transient. Therefore, there is less variability evident in these light curves, meaning that, with \texttt{dwf-postpipe} these light curves will be less likely than with \texttt{photpipe} to be identified as interesting in searches for transient phenomena.  Conversely, in the limit where $F_t >> F_g$ the two pipelines measure similar variability for the same transient.  This is expected given that, in this regime, the background flux, which is subtracted by \texttt{photpipe}, is negligible compared to that of the injected transient.

\section{SEARCH FOR VARIABLE PHENOMENA}\label{sec:variables}

Whilst DWF's observing strategy is targeted toward the detection of extragalactic fast transients, the depth and cadence of the DECam images, as well as other wavelength data, provide a unique dataset for the detection and characterisation of variable stars, including events $\sim$2--3 magnitudes fainter than current optical catalogues. Typically, surveys like the Dark Energy Survey fold many nights of observations over the course of years to identify variable stars \cite[e.g.][]{2021ApJ...911..109S}. DWF's high cadence allows for the identification of short period variables (e.g. RR Lyrae, ZZ ceti, contact binaries and delta Scuti variable stars) within a single night's observations. This temporal resolution and 6 consecutive night observing strategy, coupled with the depth afforded by DECam, results in large volumes of the Milky Way being available for the characterisation of these variables.  

Beginning in 2025, Rubin's Wide, Fast, Deep survey (WFD) will explore the transient and variable sky at an unprecedented depth and sky coverage.  In addition, with WFD's $\sim$2--5 day cadence, it may take months to properly characterise short period variables.  Our identified catalogue of stars, can be cross-matched with Rubin detections. In particular, due to our magnitude limits, we identify faint stars that can improve the target identification for variable star analyses and for reducing contaminants in extragalactic searches with Rubin brokers such as Fink \citep{2021MNRAS.501.3272M}.

Here, we present searches for variable sources in two fields, NGC Section 5.1 and CDFS, Section 5.2. In particular, The Chandra Deep Field South (CDFS) is a legacy deep field, which was observed as part of DWF field for 9 nights in total (see Table \ref{tab:fields}).  It will also be one of Rubin's deep drilling fields. DWF provides a deep, high cadence survey of CDFS across seven years. These observations could both provide supplementary data to characterise objects of interest identified with Rubin and as a filter to remove contaminants.  In Section \ref{sec:class_variables}, we classify the objects found in this search.

\subsection{Periodic sources in NGC 6101}

\begin{figure*}
    \centering
    \includegraphics[width=\textwidth]{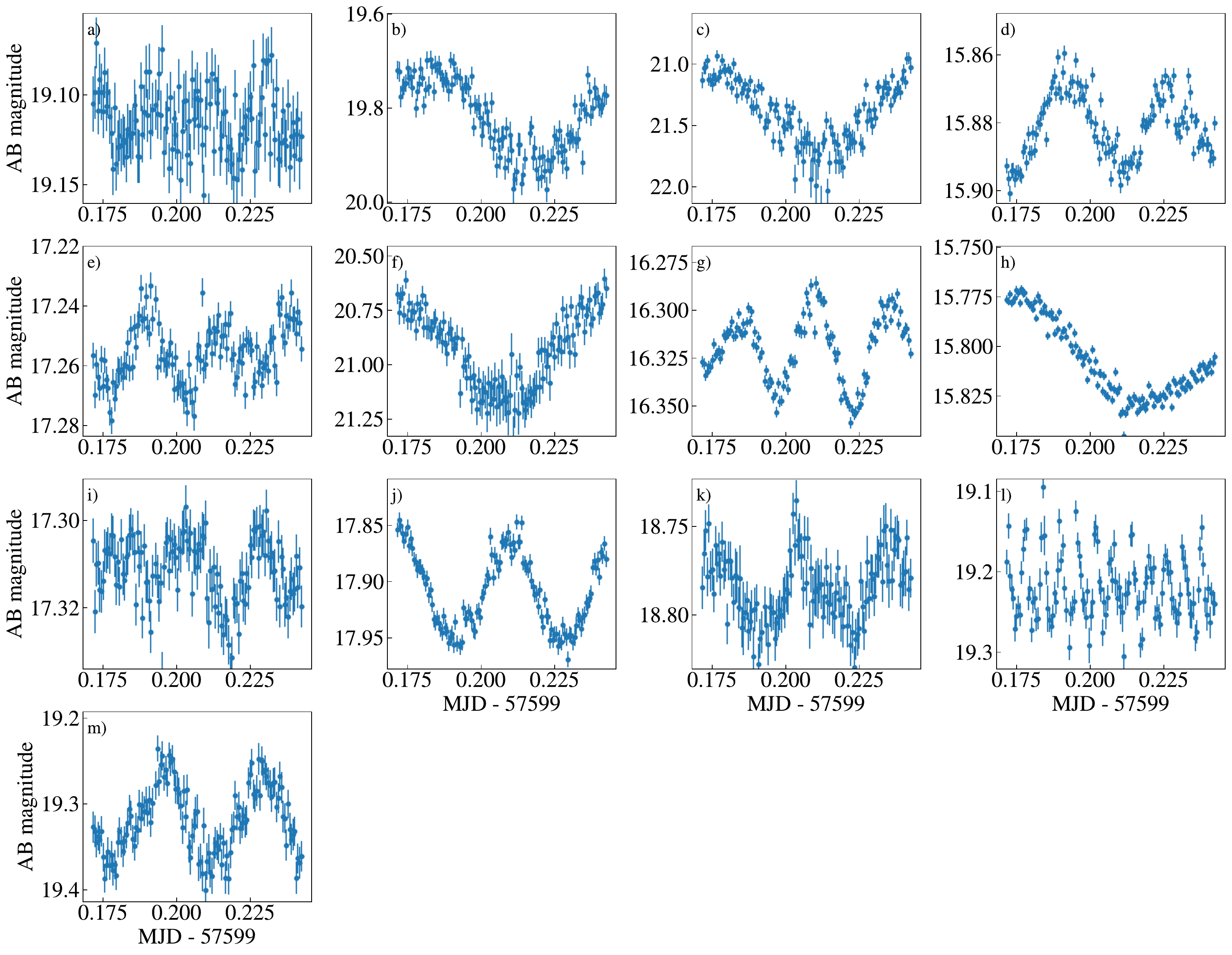}
    \caption{The light curves of uncatalogued periodic sources identified from the DWF data from the field NGC 6101 on the night of 30 July 2016 UTC.}
    \label{fig:variables}
\end{figure*}

We search a single field-night, NGC 6101 on 30 July 2016 UTC, for uncatalogued, variable stars with periods $<2$\,hr. NGC 6101 is the field with the lowest Galactic latitude and would therefore likely yield the highest density of variable stars. We generate a Lomb-Scargle periodogram \citep{1976Ap&SS..39..447L,1982ApJ...263..835S,1989ApJ...338..277P} for each source observed over the course of the night with a magnitude error $<0.1$. We cut sources that are present in variable catalogues \textit{Gaia} DR3 \citep{Gaia_DR3} and the ASAS-SN catalogue of variable stars \citep{ASAS-SN_Variables}. Sources are extracted if they have a peak frequency with a significance of $>5\sigma$ above the background. Out of the 33 sources that satisfy this cut, 13 were genuine newly discovered, periodic, astrophysical sources. The light curves of these 13 uncatalogued variable sources are shown in Fig. \ref{fig:variables} and their properties are summarised in Table \ref{tab:var}.   

\begin{table*}
 \centering
 \caption{Properties of the periodic sources found in the DWF observations of NGC 6101.  Sources without significant parallax values do not have an associated distance or luminosity measurement.}
 \label{tab:var}
 \begin{tabular*}{0.93\textwidth}{@{}l@{\hspace*{17pt}}l@{\hspace*{17pt}}l@{\hspace*{17pt}}l@{\hspace*{17pt}}l@{\hspace*{17pt}}l@{\hspace*{17pt}}l@{\hspace*{17pt}}}
  \hline
Candidate & Coordinates & Period & Distance & $m_G$ & $M_G$ & Suspected class \\
& & min & pc & AB magnitude & AB magnitude & \\
\hline
a) & 16:27:02 -72:16:07 & $27 \pm 4$ & -- & $19.138 \pm 0.004$ & -- & Pulsator \\
b) & 16:16:47 -72:23:05 & $78 \pm 3$ & $1259 \pm 183$ & $18.203 \pm 0.003$ & $7.7 \pm 0.3$ & Eclipsing Binary \\
c) & 16:35:43 -72:33:50 & $92 \pm 7$ & -- & $20.79 \pm 0.02$ & -- & Pulsator \\
d) & 16:24:45 -72:57:47 & $51 \pm 1$ & $4926 \pm 823$ & $15.900 \pm 0.003$ & $2.4 \pm 0.4$ & Pulsator \\
e) & 16:30:28 -72:51:25 & $35.1 \pm 0.7$ & -- & $17.298 \pm 0.003$ & -- & Pulsator \\
f) & 16:13:47 -73:05:56 & $102 \pm 19$ & -- & $20.23 \pm 0.01$ & -- &  Eclipsing Binary \\
g) & 16:17:15 -73:05:33 & $36.3 \pm 0.2$& $13928 \pm 8748$ & $16.429 \pm 0.003$ & $0.7 \pm 1.4$ & Pulsator \\
h) & 16:39:27 -73:01:42 & $92 \pm 4$ & $6305 \pm 1157$ & $15.745 \pm 0.003$ & $1.7 \pm 0.4$ & Pulsator \\
i) & 16:15:20 -73:15:12 & $33.9 \pm 0.8$ & $6702 \pm 3064$ & $17.320 \pm 0.004$ & $3.2 \pm 1.0$ & Pulsator \\
j) & 16:13:09 -73:21:41 & $54 \pm 1$ & -- & $17.911 \pm 0.003$ & -- & Pulsator \\
k) & 16:30:02 -73:34:47 & $42 \pm 1$ & $4613 \pm 3685$ & $18.834 \pm 0.003$ & $5.5 \pm 1.7$ & Pulsator \\
l) & 16:28:36 -73:31:39 & $8.62 \pm 0.03$ & $939 \pm 268$ & $19.469 \pm 0.004$ & $9.6 \pm 0.6$ & ZZ ceti \\
m) & 16:19:55 -73:45:21 & $46.2 \pm 0.9$ & -- & $19.265 \pm 0.004$ & -- & Pulsator \\
  \hline
 \end{tabular*}
\end{table*}

\subsection{Variable sources in CDFS}

We search for fast-varying objects in the CDFS field observed for nine field-nights during four DWF operational runs. For the purposes of this search, we define a variable object as one that has an intra-night light curve which satisfies $\eta^{-1} > 1$. We restrict this search only to sources that vary on fast, intra-night timescales. We also remove sources that are present in variable star catalogues, including \textit{Gaia} DR3 \citep{Gaia_DR3}, the ASAS-SN catalogue of variable stars \citep{ASAS-SN_Variables} and the catalogue for RR Lyrae variable stars in DES Y6 \citep{2021ApJ...911..109S}. With these cuts, we were left with $\sim$3000 objects. After visual inspection of these candidates, all but two of them were assessed to be spurious. The spurious objects were either sources displaying variability due to their proximity to the edge of a CCD, electronic artefacts or their proximity to diffraction spikes from a nearby bright star.

\begin{figure*}
    \centering
    \includegraphics[width=\textwidth]{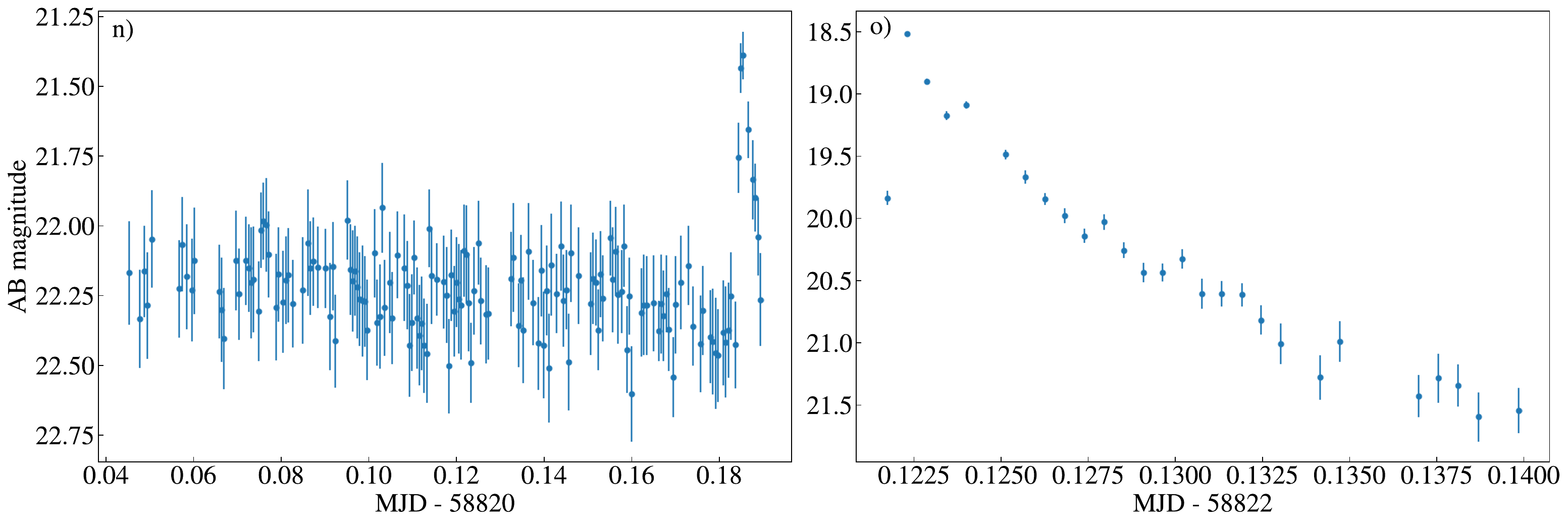}
    \caption{Light curves of variable phenomena (candidates \textit{n} and \textit{o}) found by searching all DWF observations of CDFS.}
    \label{fig:CDFS_flares}
\end{figure*}

The remaining two candidates are summarised in Table \ref{tab:flares} and their light curves are shown in Fig. \ref{fig:CDFS_flares}. Neither of these candidates display variability on any nights other than those plotted in Fig. \ref{fig:CDFS_flares} and display a fast rise and slower decay which indicates that they are likely eruptive, flare-like events.  

\begin{table*}
 \centering
 \caption{Summary of variable phenomena found by searching the CDFS field during four DWF operational runs.}
 \label{tab:flares}
 \begin{tabular*}{0.8\textwidth}{@{}l@{\hspace*{17pt}}l@{\hspace*{17pt}}l@{\hspace*{17pt}}l@{\hspace*{17pt}}l@{\hspace*{17pt}}}
 \hline
 Candidate & Coordinates & Peak time & Peak $g$-band brightness & Quiescent $g$-band brightness\\
 & & UTC & AB magnitude & AB magnitude\\
 \hline
 n) & 03:31:40 -29:03:29 & 2019-12-03T04:26:58 & $21.39 \pm 0.09$ & $22.23\pm0.03$ \\
 o) & 03:31:34 -27:45:04 & 2019-12-05T02:56:07 & $18.52 \pm 0.02$ & $22.15\pm0.03$ \\
 \hline
\end{tabular*}
\end{table*}

\subsection{Classification of variable sources}\label{sec:class_variables}

\begin{figure}
    \includegraphics[width=\columnwidth]{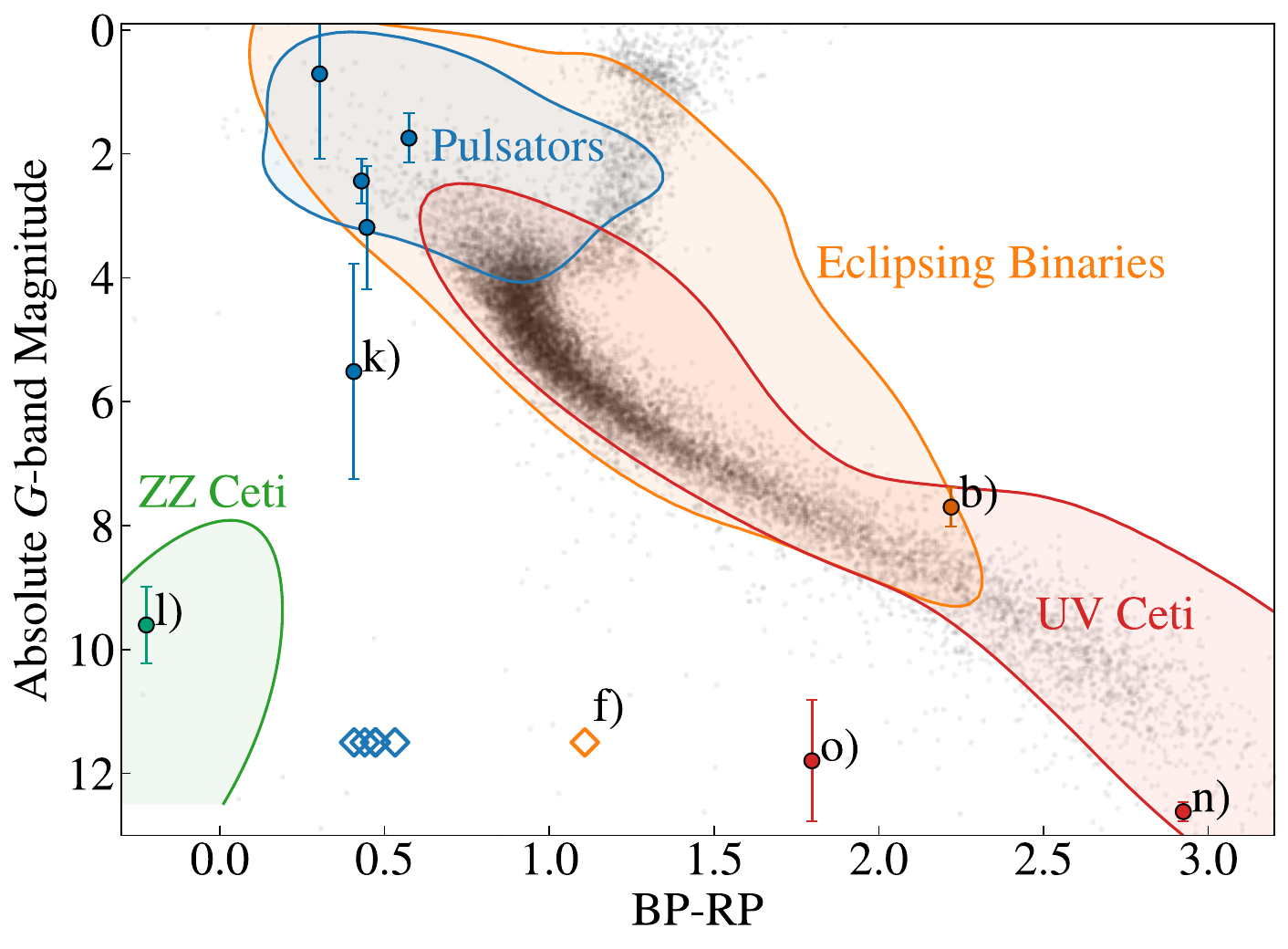}
    \caption{Absolute $G$-band magnitude and BP-RP values of the candidates present in Figs. \ref{fig:variables} and \ref{fig:CDFS_flares}, obtained from Gaia DR3 \citep{Gaia_DR3}.  Sources without significant parallax values do not have an associated distance or absolute magnitude measurement and are denoted with diamonds.  The black points denote a random sample of Gaia DR3 sources with significant parallax measurements.   The contour lines denote the regions in parameter space where different variable star types inhabit.  ZZ ceti, eclipsing binaries, UV ceti and pulsating variables are plotted.  Pulsating variables include $\delta$ Scuti, RR Lyrae and SX Pheonicis type stars.  The contours are constructed using sources from The International Variable Star Index \cite{2006SASS...25...47W}, cross-matched with Gaia DR3.}
    \label{fig:HR_diagram}
\end{figure}

To help classify the periodic sources found in our NGC 6101 field search, we cross-match to Gaia DR3 \citep{Gaia_DR3} to obtain both parallax and Gaia satellite's Blue and Red Photometer (BP/RP) measurements \citep{2023A&A...674A...2D}. This information allows us to construct a colour-magnitude diagram to help determine a spectral classification. The candidates that have coincident detections in Gaia DR3 are presented in Figs. \ref{fig:variables} and \ref{fig:CDFS_flares}. However, whilst all candidates have BP/RP values, a subset either do not have parallax measurements or possess errors larger than the parallax value itself. Despite this, we are able to provide a tentative classification of all sources found in Fig. \ref{sec:variables}. From their location in Fig. \ref{fig:HR_diagram} and the plateau-like phases in their light curves, we classify candidates \textit{b} and \textit{f} as eclipsing binaries. Candidate \textit{l} is likely a ZZ Ceti variable star \citep[e.g.][]{2008PASP..120.1043F,2010A&ARv..18..471A}, due to its location in Fig. \ref{fig:HR_diagram} and its short period of $8.62\pm0.03$ minutes.  The remaining candidates are likely a combination of RR Lyrae, $\delta$ Scuti and SX Phoenicis type stars.

For the variable sources found in CDFS, we classify candidate \textit{n} as a UV Ceti due to its location in Fig. \ref{fig:HR_diagram}. These are often a low-mass, usually M-class, stars characterised by sporadic flares. However, candidate \textit{o} is subluminous for this population. This could be due to the star being a subdwarf or possessing a blue, faint companion star such as a white dwarf. It also has an X-ray counterpart in the Chandra source catalogue \citep[CXO J033133.7-274505;][]{2010ApJS..189...37E} and the XMM Newton serendipitous source catalogue \citep[4XMM J033133.7-274505;][]{2020A&A...641A.136W}.

Given the high galactic latitude ($-54.93^\circ$) and sparsity of CDFS, it is expected that a smaller number of uncatalogued variables would be identified compared to NGC 6101. CDFS is also a well-explored region of sky, meaning it is well populated with public variable star catalogues. From this, we conclude that variable star catalogues will be effective in removing contaminants for well-explored extragalactic fields like CDFS for $g<22.2$ AB mag.

\section{DISCUSSION}\label{sec:discussion}

\subsection{Broader applications and impact}

DWF's first optical data release provides a dataset which probes a parameter space poorly characterised by most other transient surveys. We have shown that \texttt{dwf-postpipe} provides a set of data products which is effective in extracting science from this dataset, outperforming a difference imaging approach in some scenarios and providing light curves for all sources in the fields. The science-ready data products, available on AAO Data Central, will allow for simple exploration and querying of this vast dataset.

Some of these data have been mined using various methods for extragalactic fast transients \citep[e.g.][]{2020MNRAS.491.5852A,2024MNRAS.531.4836F} and stellar flares \citep[e.g.][]{2020MNRAS.498.3077W,2021MNRAS.506.2089W}, Section \ref{sec:variables} and \ref{sec:class_variables} has demonstrated that there is potential for studying variable star populations and different transient types and durations via other techniques. The exploration of one night's observations of NGC 6101 here has produced a sample of 13 uncatalogued, fast-evolving, stellar pulsators, with the entire DWF dataset harbouring more. We leave a comprehensive search to future work but this sample would likely provide a number of well understood pulsators such as $\delta$ scuti, SX Phoenicus and RR Lyrae type stars. Photometric observations of these stars provide a means of studying stellar interiors via their oscillations \citep{1995A&A...293...87K}. DWF DR1 probes large Galactic volumes, which may enable population studies of white dwarf pulsators.

While DWF provides limited sky coverage compared to Rubin, DWF DR1 can complement Rubin and its brokers by providing high-cadence coverage sources of interest identified in the Wide, Fast, Deep survey. It will provide characterisation of short period variable stars, sampling their full periods. In addition, in the future, we plan to release the full, multi-wavelength DWF dataset on AAO Data Central including simultaneous radio, high energy and optical data from Subaru/Hyper Suprime-Cam and KMTNeT. We also plan to provide nightly stacked difference imaging, which will reach depths of $g\sim25$--$26$ AB mag and will be coupled with multi-band imaging in $r$ and $i$-band to constrain colour evolution.

\subsection{Future work}

In recent years, low read noise, scientific-grade Complementary Metal-Oxide-Semiconductor (CMOS) detectors have increasingly been used in astronomical facilities. Their fast readout times result in a cadence unachievable with CCD detectors without sacrificing duty cycle. While CMOS detectors and electron-multiplying CCDs (EMCCDs) on major facilities \citep[e.g.][]{2014MNRAS.444.4009D,2006MNRAS.372..151O} have small fields of view, making transient search programs difficult, there are upcoming telescope arrays with wide fields of view. The Large Array Survey Telescope \citep[][]{2023PASP..135h5002B} and the upcoming Argus Array \citep{2022PASP..134c5003L} utilise an array of off-the-shelf CMOS detectors to construct observatories with sensitivities equivalent to several meter class telescopes. The Argus Array's instantaneous sky coverage is 7916 square degrees, reaching a $r\sim20$ AB mag depth, similar to ZTF, in a few minutes and a $r\sim22$ AB mag depth, similar to DWF, in one hour. Facilities like these will be the next steps in characterising the dynamic optical sky.

\section{SUMMARY}

The transient and variable optical sky is poorly characterised at seconds-to-hours timescales. Traditional, current generation, transient surveys such as Zwicky Transient Facility's main survey and the Dark Energy Survey have cadences in excess of a day. Other high cadence surveys such as Evryscope and TESS possess a very wide field-of-view but are comparatively shallow. The Deeper, Wider, Faster programme (DWF), utilising the Dark Energy Camera (DECam), probes a unique parameter space between these two observing strategies, achieving deep ($g\sim22.2$), minute-cadence observations with a three square degree sky coverage per pointing.

In this work, we present DWF's first data release of optical data products of the minute-cadence DECam observations collected during the 14 DWF observing runs to-date (excludes Subaru Hyper Suprime-Cam and KMTNet observations). The dataset includes 112 DECam pointings, each with 0.5--3\,hr of continuous, minute-cadence, observations and a cumulative time on-sky time of 166\,hr.

The data products were produced from a novel data processing pipeline, \texttt{dwf-postpipe}.  For each field, for each night, this pipeline takes in calibrated images from the DECam community pipeline and outputs lightcurves for every source in the field.  It does this in three steps; identifying all transient and quiescent sources visible throughout each night's images, aligning these images to a `detection image' and finally using this detection image to extract time-resolved photometry.  The result is a rich set of observations of transient and variable phenomena that are publicly available on AAO Data Central.

These observations reach a median 5$\sigma$ depth of 22.2 AB mag and a median seeing FWHM of 1.35\arcsec.  We assess the data quality by comparing our photometric measurements of quiescent sources to that of DELVE. We found our observations, calibrated with \textit{SkyMapper} DR4, have a consistent, systematic, $\sim$0.15 magnitude offset from DELVE. Fake sources were injected directly into images, extracted, and light curves generated to compare the efficiency and recovery of transients with a typical difference imaging approach for DECam, \texttt{photpipe}. We find similar performance with the two approaches to 22\,AB magnitude and an efficiency of $97.24^{+0.7}_{-1.0}$ percent for \texttt{dwf-postpipe} and $96.14^{+0.9}_{-1.1}$ percent for \texttt{photpipe}.  \texttt{dwf-postpipe} recovers a greater number of sub-threshold transients $g>22$ with an efficiency of $63.9^{+2.7}_{-2.8}$ percent compared to \texttt{photpipe}'s $29.3^{+2.7}_{-2.6}$ percent. We attribute this to the loss of sensitivity imparted by image subtraction. However, we note that as \texttt{dwf-postpipe} measures photometry directly on the science images, transients with a comparatively bright background galaxy are difficult to identify as the variability in the transient signal competes with the noise induced by the underlying host galaxy emission.

Whilst DWF is a unique survey for identifying fast transients, it also provides a unique avenue to study variable stars.  In a search for uncatalogued variable stars in one night of DWF observations of one field, NGC 6101, we find 13 such sources. Of these sources ten are pulsating variables, two are eclipsing binaries and one is a ZZ ceti star.

DWF's observations of Chandra Deep Field South, a Rubin deep-drilling field, provides a dataset to flag rapidly evolving and faint variable sources which may take many months/years before they are identified and characterised as a result of Rubin's observational cadence. We search nine nights of DWF observations of CDFS for uncatalogued faint variable sources and identify two flares from likely UV ceti type stars. From this, we conclude that in CDFS and, more generally, deep, extragalactic, legacy fields existing optical variable star catalogues are effective in removing variable star contaminants to $g<22.2$ AB mag. In the Rubin-era, these catalogues will play a crucial role in filtering in the deep-drilling fields which will allow for more efficient detection of novel, extragalactic transients \citep[e.g.][]{2025arXiv250722864F}.

\section{DATA AVAILABILITY}

The photometry and image cutouts can be queried on AAO Data Central (\url{https://doi.org/10.57891/gy62-bj96}) and includes a range of summary statistics and catalogue cross-matches.  Raw and calibrated images are available on the NOIRLab Astro Data Archive under program numbers 2015B-0607, 2016A-0095, 2017A-0909, 2019A-0911, 2018A-0137, 2019B-1012 and 2020B-0253. Observations that are still within the 18-month proprietary period and all other code and data underlying this work will be shared upon reasonable request to the authors.

\begin{acknowledgement}

We thank Courtney Crawford and Ben Montet for valuable discussions regarding variable star candidate identification and classification.

We also thank the anonymous referee for their insightful comments.

J.C. acknowledges funding by the Australian Research Council Discovery Project, DP200102102.

A.M. is supported by the Australian Research Council DE230100055.

Parts of this research were conducted by the Australian Research Council Centre of Excellence for Gravitational Wave Discovery (OzGrav), through project numbers CE170100004 and CE230100016.

This paper includes data that has been provided by the AAO Data Central Science Platform (datacentral.org.au) and makes use of services and code that have been provided by the AAO Data Central Science Platform.

This project used data obtained with the Dark Energy Camera (DECam), which was constructed by the Dark Energy Survey (DES) collaboration. Funding for the DES Projects has been provided by the US Department of Energy, the U.S. National Science Foundation, the Ministry of Science and Education of Spain, the Science and Technology Facilities Council of the United Kingdom, the Higher Education Funding Council for England, the National Center for Supercomputing Applications at the University of Illinois at Urbana-Champaign, the Kavli Institute for Cosmological Physics at the University of Chicago, Center for Cosmology and Astro-Particle Physics at the Ohio State University, the Mitchell Institute for Fundamental Physics and Astronomy at Texas A\&M University, Financiadora de Estudos e Projetos, Fundação Carlos Chagas Filho de Amparo à Pesquisa do Estado do Rio de Janeiro, Conselho Nacional de Desenvolvimento Científico e Tecnológico and the Ministério da Ciência, Tecnologia e Inovação, the Deutsche Forschungsgemeinschaft and the Collaborating Institutions in the Dark Energy Survey.

The Collaborating Institutions are Argonne National Laboratory, the University of California at Santa Cruz, the University of Cambridge, Centro de Investigaciones Enérgeticas, Medioambientales y Tecnológicas–Madrid, the University of Chicago, University College London, the DES-Brazil Consortium, the University of Edinburgh, the Eidgenössische Technische Hochschule (ETH) Zürich, Fermi National Accelerator Laboratory, the University of Illinois at Urbana-Champaign, the Institut de Ciències de l’Espai (IEEC/CSIC), the Institut de Física d’Altes Energies, Lawrence Berkeley National Laboratory, the Ludwig-Maximilians Universität München and the associated Excellence Cluster Universe, the University of Michigan, NSF NOIRLab, the University of Nottingham, the Ohio State University, the OzDES Membership Consortium, the University of Pennsylvania, the University of Portsmouth, SLAC National Accelerator Laboratory, Stanford University, the University of Sussex, and Texas A\&M University.

Based on observations at NSF Cerro Tololo Inter-American Observatory, NSF NOIRLab (NOIRLab Prop. ID 2020B-0253; PI: J. Cooke), which is managed by the Association of Universities for Research in Astronomy (AURA) under a cooperative agreement with the U.S. National Science Foundation.

This research made use of \texttt{matplotlib}, a Python library for publication quality graphics \citep{matplotlib}, \texttt{SciPy} \citep{scipy}, \texttt{Astropy}, a community-developed core Python package for Astronomy \citep{astropy1,astropy2} and \texttt{scikit-learn}   \citep{sklearn}.

\end{acknowledgement}


\bibliography{ref}

\appendix

\section{Error Correction Calculation}\label{sec:error_correction}

Performing a Gaussian convolution on an image with a FWHM of $\sigma_c$, neighbouring pixels become correlated.  This reduces the effective number of independent pixels, $N_A$, compared to the actual number of pixels in the convolved image, $N$, summed to calculate the flux, $f$
\begin{equation}
    N_A = \frac{N}{2\pi\sigma_c}
\end{equation}
The background variance, $\sigma_{bkg}^2$, is calculated globally from all pixels in the image, and is conserved when the image is convolved
\begin{align}
    \sigma_{\mathrm{bkg/pix}}^2 &= \sigma_{\mathrm{bkg/pix},c}^2 \\
    \sigma_{\mathrm{bkg}}^2     &= N_A\sigma_{\mathrm{bkg/pix},c}^2 = \frac{1}{2\pi\sigma_c}\sigma_{\mathrm{bkg},c}^2
\end{align}
The root-mean-squared error for a given source, $\sigma_{\mathrm{src}}^2$ is given by the sum of the variance due the shot noise, $\sigma_{\mathrm{shot}}^2$ and background noise, $\sigma_{\mathrm{bkg}}^2$
\begin{align}    
    \sigma_{\mathrm{src}}^2 &= \sigma_{\mathrm{shot}}^2 + \sigma_{\mathrm{bkg}}^2 \\
                            &= \frac{1}{g}\sum_i^{N_A}f_i + \sigma_{\mathrm{bkg}}^2
\end{align}
where $g$ is the gain.  Assuming the flux for a given source in the convolved image is conserved,
\begin{align}    
    \sigma_{\mathrm{src}}^2 &= \frac{1}{2\pi\sigma_c}\frac{1}{g}\sum_i^{N}f_{i,c} + \sigma_{\mathrm{bkg}}^2 \\
                            &= 2\pi\sigma_c(\sigma_{\mathrm{shot},c}^2 + \sigma_{\mathrm{src},c}^2) \\
                            &= \alpha(\sigma_{\mathrm{shot},c}^2 + \sigma_{\mathrm{src},c}^2)\label{eq:err_correction}
\end{align}
Assuming two consecutive images have similar seeing and depth, we can calculate $\alpha$ by comparing the median difference in flux $\Bar{f}_{\mathrm{diff},j}$ between consecutive images to the median flux error $\Bar{\sigma}_{\mathrm{src,c,j}}$, for infinitesimal flux bins which have a median flux, $\Bar{f}_j$. Using Equation \ref{eq:err_correction},
\begin{equation}
    \alpha = \frac{\Bar{f}_{\mathrm{diff},j}^2}{\Bar{f}^2_j}
\end{equation}

\section{Catalogue Schema}\label{sec:schema}

\begin{table*}
\centering
\caption{Columns in the \texttt{mastercatalogue} catalogue}
\label{tab:mastercatalogue}
\begin{tabular*}{0.62\textwidth}{lll}
\hline
Name & Description & Unit \\
\hline
\texttt{candidate\_id} & Unique identifier for each transient or variable candidate & - \\
\texttt{ccd} & Charge-coupled device identifier on the Dark Energy Camera & - \\
\texttt{night} & Observing night identifier (YYYYMMDD format) & - \\
\texttt{ra} & Right Ascension of the candidate & deg \\
\texttt{dec} & Declination of the candidate & deg \\
\texttt{mjd\_start} & Modified Julian Date at the start of observations & days \\
\texttt{mjd\_end} & Modified Julian Date at the end of observations & days \\
\texttt{mjd\_mid} & Modified Julian Date at the midpoint of observations & days \\
\texttt{med\_uplim} & Median upper limit magnitude of observations & AB mag \\
\texttt{var} & von Neumann statistic of light curve & - \\
\texttt{detections} & Number of detections of the source & - \\
\texttt{max\_mag} & Faintest observed g-band magnitude (highest numerical value) & AB mag \\
\texttt{min\_mag} & Brightest observed g-band magnitude (lowest numerical value) & AB mag \\
\texttt{stdmag} & Standard deviation of the observed magnitudes & AB mag \\
\texttt{mederr} & Median photometric uncertainty of the detections & AB mag \\
\texttt{medmag} & Median observed g-band magnitude & AB mag \\
\hline
\end{tabular*}
\end{table*}

\begin{table*}
\centering
\caption{Columns in the \texttt{lightcurve} catalogue}
\label{tab:lightcurve}
\begin{tabular*}{0.62\textwidth}{lll}
\hline
Name & Description & Unit \\
\hline
\texttt{candidate\_id} & Unique identifier for each transient or variable candidate & - \\
\texttt{ccd} & Charge-coupled device identifier on the Dark Energy Camera & - \\
\texttt{night} & Observing night identifier (YYYYMMDD format) & - \\
\texttt{mjd} & Modified Julian Date of the observation & days \\
\texttt{x\_pix} & X coordinate of the source centroid in image pixels & pixel \\
\texttt{y\_pix} & Y coordinate of the source centroid in image pixels & pixel \\
\texttt{ra} & Right Ascension of the source & deg \\
\texttt{dec} & Declination of the source & deg \\
\texttt{flux} & Measured source flux in g-band & $\mu$Jy \\
\texttt{flux\_err} & Uncertainty in measured flux & $\mu$Jy \\
\texttt{mag} & Measured g-band magnitude & AB mag \\
\texttt{mag\_err} & Uncertainty in measured magnitude & AB mag \\
\texttt{fwhm\_pix} & Full Width at Half Maximum of the source in image pixels & pixel \\
\texttt{fwhm} & Full Width at Half Maximum of the source in arcseconds & \arcsec \\
\texttt{ellipticity} & Ellipticity of the source profile (1 - B/A) & - \\
\texttt{limiting\_mag} & Limiting magnitude (5-sigma) for the observation & AB mag \\
\texttt{filename} & Name of the image file containing the measurement & - \\
\hline
\end{tabular*}
\end{table*}

\end{document}